\documentclass[useAMS,usenatbib]{mn2e}
\usepackage{graphicx}

\title[Constraints on opacities from complex asteroseismology of B-type pulsators]
      {Constraints on opacities from complex asteroseismology of B-type pulsators: the $\beta$ Cephei star $\theta$ Ophiuchi}

   \author[J. Daszy\'nska-Daszkiewicz, P. Walczak]
   {J. Daszy\'nska-Daszkiewicz\thanks{E-mail:daszynska@astro.uni.wroc.pl},
    P. Walczak\thanks{E-mail:walczak@astro.uni.wroc.pl}\\
   Instytut Astronomiczny, Uniwersytet Wroc{\l}awski,
   ul. Kopernika 11, 51-622 Wroc{\l}aw, Poland\\
    }
    \begin{document}

   \date{Received ...; accepted ...:in orginal form ...}

   \pagerange{\pageref{firstpage}--\pageref{lastpage}} \pubyear{2009}

   \maketitle

\label{firstpage}

\begin{abstract}
We present results of a {\bf comprehensive} asteroseismic modelling of the $\beta$
Cephei variable $\theta$ Ophiuchi. {\bf We call these studies {\it complex asteroseismology}
because our goal is to reproduce both pulsational frequencies as well as corresponding
values of a complex, nonadiabatic parameter, $f$, defined by the radiative flux perturbation.}
To this end, we apply the method of simultaneous determination of the spherical harmonic degree,
$\ell$, of excited pulsational mode and the corresponding
nonadiabatic $f$ parameter from combined multicolour photometry and
radial velocity data. Using both the OP and OPAL opacity data, we
find a family of seismic models which reproduce the radial and
dipole centroid mode frequencies, as well as the $f$ parameter
associated with the radial mode. Adding the nonadiabatic
parameter to seismic modelling of the B-type main sequence pulsators
yields very strong constraints on stellar opacities. In particular,
only with one source of opacities it is possible to agree the
empirical values of $f$ with their theoretical counterparts. Our
results for $\theta$ Oph point substantially to preference for the
OPAL data.
\end{abstract}

\begin{keywords}
stars: $\beta$ Cephei variables --
stars: individual: $\theta$ Ophiuchi --
stars: pulsation --
stars: opacities
\end{keywords}

\section{Introduction}
During the last two decades, $\beta$ Cephei stars became attractive
targets for asteroseismic studies. This has started with the
pioneering papers by Dziembowski \& Jerzykiewicz on 16 Lac and 12
Lac (Dziembowski \& Jerzykiewicz, 1996 and 1999, respectively). The
analysis of multiplets in 16 Lac showed that rotation rate of this
star has to increase inward. The seismic modelling of B-type
pulsators have been revived with the paper by Aerts, Toul,
Daszy\'nska et al. (2003) on V836 Cen, in which they also found a
non-rigid rotation and, for the first time, an evidence for
overshooting from a convective core in a massive main sequence star.
Up to now, the most intensively studied $\beta$ Cephei stars have
been $\nu$ Eri and 12 Lac which were subjects of multi-site
photometric and spectroscopic campaigns (Handler et al. 2004, Aerts
et al. 2004, Jerzykiewicz et al. 2005, Handler et al. 2006). The
seismic modelling of $\nu$ Eri was done by Pamyatnykh, Handler \&
Dziembowski (2004) and Ausseloos et al. (2004). Stronger constraints
on stellar structure can be obtained from asteroseismology of the
hybrid B-type pulsators in which two types of modes are excited
simultaneously, i.e., low-order pressure and gravity modes (typical
for $\beta$ Cephei stars) and high-order gravity modes (typical for
Slowly Pulsating B-type stars). Such studies were done recently by
Dziembowski \& Pamyatnykh (2008) for $\nu$ Eri and 12 Lac.
Interesting results are coming up for another hybrid pulsating star
$\gamma$ Peg (Handler et al. 2009).

Until now, asteroseismic studies have relied on fitting only
pulsation frequencies. The ultimate goal has been to find models
which reproduce observed frequencies by changing various parameters
of model and theory, provided that the successful mode
identification was obtained beforehand. However oscillation
frequencies are determined mainly by the star's interior, thus they
are rather weakly sensitive to the subphotospheric layers. A few
years ago, Daszy\'nska-Daszkiewicz, Dziembowski \& Pamyatnykh (2003
and 2005a, hereafter DD03 and DD05) introduced a new asteroseismic
probe, defined as the ratio of the amplitude of the bolometric flux
variations to the radial displacement at the photosphere level. This
is so called $f$ parameter; it is determined in the pulsation
driving region.

The $f$ parameter exhibits a strong dependence on the global stellar
parameters, chemical composition, opacities and subphotospheric
convection. Therefore, a comparison of empirical values of the $f$
parameter with their theoretical counterparts can yield valuable and
unique information on stellar structure and evolution. The
theoretical values of $f$ result from linear nonadiabatic
computations of stellar pulsation. To explore the asteroseismic
potential of the $f$ parameter, DD03 proposed the method of
simultaneous determination of the spherical harmonic degree, $\ell$,
and the nonadiabatic $f$ parameter from amplitudes and phases of
photometric and radial velocity variations. In this way we can also
circumvent uncertainties in mode identification resulting from
pulsational input. Using this method one can derive also an estimate
of intrinsic amplitude of pulsational modes.

This method has been successfully applied to $\delta$ Scuti and
$\beta$ Cep stars. In the case of $\delta$ Scuti variables,
stringent constraints on the subphotospheric convection were
obtained (DD03 and Daszy\'nska-Daszkiewicz et al. 2005b). For the
$\delta$ Scuti objects studied by these authors, convective energy
transport appeared to be inefficient. For $\beta$ Cephei stars, DD05
have shown that the value of $f$ is very sensitive to the
metallicity parameter, $Z$, and to the adopted opacity tables.
Moreover, in the case of B-type pulsators, an unambiguous solution
exists only if one combines photometry and radial velocity data.

In this paper we present an asteroseismic study which aims at
parallel fitting of the pulsational frequencies and corresponding
values of {\bf the complex $f$ parameter}, taking into account instability conditions.
We will call such studies {\it complex asteroseismology}. For our analysis
we chose the $\beta$ Cep star $\theta$ Oph for which time series
photometric and spectroscopic data were acquired by Handler,
Shobbrook \& Mokgwetsi (2005, hereafter HSM05) and Briquet et al.
(2005, hereafter B05), respectively. Seismic modelling of $\theta$
Oph was recently done by Briquet et al. (2007) whose computations
were based only on the OP tables. Here we shall use both OP and OPAL
opacities and incorporate the $f$ parameter in the seismic model
survey.

The paper is composed as follows. In Section 2 we view the basic
characteristics of $\theta$ Oph. Section 3 is devoted to mode
identification of the excited modes using two approaches.
Constraints on intrinsic mode amplitude are given in Section 4.
Results of complex asteroseismic modelling, adopting the OP and OPAL
tables, are given in Section 5. A Summary ends the paper.

\section{The pulsating star $\theta$ Ophiuchi}
$\theta$ Oph is the $\beta$ Cep pulsator with the B2IV spectral type
and the brightness of $V=3.25$ mag. It is also a triple system, with
the speckle component of the B5 spectral type (McAlister et al.
1993, HSM05). In addition, B05 detected a low mass companion
($M<1M_\odot$) from spectroscopic observations; the orbital period
of this component is equal to 56.71 d. The main component of the
$\theta$ Oph system ($\theta$ Oph A), has been recognized as a
pulsating variable of the $\beta$ Cep type already in 1922 by
Henroteau (Henroteau 1922) on the basis of radial velocity
observations. The correct value of the period of about 0.14 d was
derived by McNamara (1957). The projected rotational velocity of
$\theta$ Oph A is about 30 km/s as measured by  Abt, Levato \&
Grosso (2002). More information on the $\theta$ Oph system can be
found in the catalog of galactic $\beta$ Cephei stars by Stankov \&
Handler (2005). A few years ago, the study of $\theta$ Oph A,
$\theta$ Oph henceforth, has been resumed thanks to dedicated
photometric and spectroscopic campaigns organized by HSM05 and B05,
respectively.

In Fig.\,1 we show the position of $\theta$ Oph in the HR diagram.
The evolutionary tracks from ZAMS to TAMS were computed with the Warsaw-New Jersey
evolutionary code adopting the OP opacities (Seaton 2005) and the
solar mixture as determined by Asplund et al. (2004, 2005, hereafter
A04). The rotational velocity at the ZAMS was assumed to be equal to
30 km/s. As $\theta$ Oph has metallicity, $Z$, lower than 0.015,
(Daszy\'nska 2001, Niemczura \& Daszy\'nska-Daszkiewicz 2005, B05),
we showed also the effect of the $Z$ parameter on the evolutionary
tracks. The observational values of effective temperature and
luminosity for $\theta$ Oph were taken from HSM05. Determination of
effective temperature from spectroscopy gives much higher values,
$\log T_{\rm eff}=4.36-4.42$. For the purpose of later discussion we
plotted also lines of constant period with the value of 0.1339 d for
the radial fundamental and first overtone mode.
\begin{figure}
\centering
\includegraphics[width=88mm,clip]{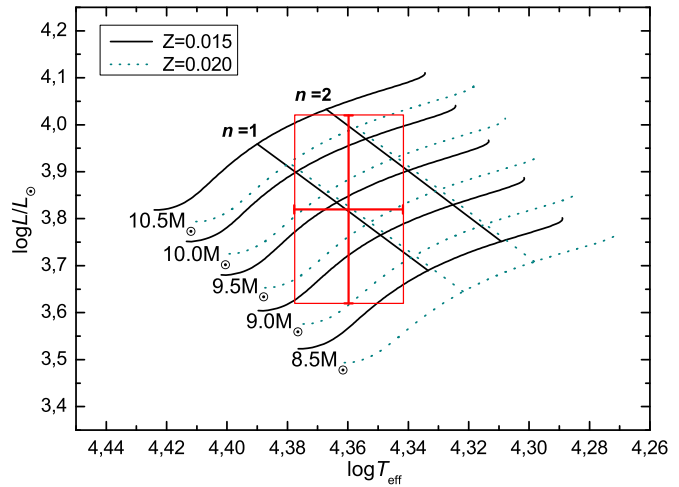}
\caption{ The observational error box of $\theta$ Ophiuchi in the HR
diagram as determined from photometry by HSM05. The evolutionary
tracks were computed with the OP-A04 opacities, assuming the
rotational velocity of $V_{\rm rot}=30$ km/s, hydrogen abundance,
$X=0.7$, two values of the metallicity parameter $Z$, and no
overshooting from the convective core. We show also lines of
constant period (0.1339 d) for the radial fundamental $(n=1)$ and
the first overtone $(n=2)$ mode. These lines will be discussed later
on.} \label{aaaaaa}
\end{figure}
\begin{figure}
\centering
\includegraphics[width=88mm,clip]{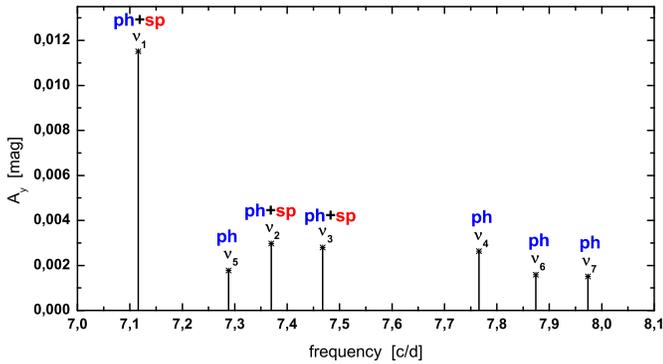}
\caption{ Oscillation spectrum of $\theta$ Ophiuchi. Seven
frequencies were extracted from photometry (HSM05), marked as ${\bf
ph}$, and three of them were found in spectroscopic variations
(B05), marked as ${\bf sp}$.} \label{aaaaaa}
\end{figure} %

The frequency analysis of time-series multicolour photometry of
$\theta$ Oph, performed by HSM05, revealed seven pulsational
frequencies in the range from 7.1160 to 7.9734 c/d. The dominant
frequency ($\nu_1=7.1160$ c/d) corresponds to the earlier detected
period (0.14 d) by McNamara (1957). Three of these seven frequncies
were identified also in spectroscopy by B05. In Fig.\,2, we show the
oscillation spectrum of $\theta$ Oph. With {\bf ph} we marked peaks
detected in photometry and with {\bf ph+sp} those three peaks
detected both in photometry and spectroscopy.

\section{Identification of pulsational modes of $\theta$ Oph}
For the time being, mode identification in main sequence pulsators
cannot be obtained from oscillation spectra themselves. Therefore,
one has to rely on other observational quantities which bring
information on the mode geometry. Such observables are, for example,
amplitudes and phases of the multicolour photometric and radial
velocity variations. Their theoretical values are computed under the
assumptions of linear oscillation and plane-parallel, temporally
static approximation for atmospheres. These assumptions are
justified in the case of main sequence pulsators. If the effects of
rotation on pulsation are ignored then the expression for the
complex amplitude of the light variations in the passband $\lambda$
can be written in the form (see, e.g., Daszy\'nska-Daszkiewicz et
al. 2002):
$${\cal A}_{\lambda}(i) = -1.086 \varepsilon Y_{\ell}^m(i,0) b_{\ell}^{\lambda}
(D_{1,\ell}^{\lambda}f+D_{2,\ell}+D_{3,\ell}^{\lambda})\eqno(1)$$
where
$$D_{1,\ell}^{\lambda} = \frac14  \frac{\partial \log ( {\cal
F}_\lambda |b_{\ell}^{\lambda}| ) } {\partial\log T_{\rm{eff}}} ,
\eqno(2a)$$
$$D_{2,\ell} = (2+\ell )(1-\ell ), \eqno(2b)$$
$$D_{3,\ell}^{\lambda}= -\left( 2+ \frac{\omega^2 R^3}{GM} \right)
 \frac{\partial \log ( {\cal F}_\lambda |b_{\ell}^{\lambda}|
) }{\partial\log g_{\rm eff}}. \eqno(2c)$$
$\varepsilon$ is the mode intrinsic amplitude, $i$ is the
inclination angle and $\ell,m$ are the spherical harmonic degree and
the azimuthal order of the mode, respectively. Symbols
$G,M,R,\omega$ have their usual meanings. The amplitudes and phases
are given by $A_{\lambda}=|{\cal A}_\lambda|$ and
$\varphi_{\lambda}=arg({\cal A}_\lambda)$, respectively.

The term $D_{1,\ell}^\lambda$ describes temperature effects,
$D_{2,\ell}$ - geometrical effects, and $D_{3,\ell}^\lambda$ - the
influence of pressure changes. The terms $D_{1,\ell}^\lambda$ and
$D_{3,\ell}^\lambda$ include the perturbation of the limb-darkening.
$b_{\ell}^{\lambda}$ is the disc-averaging factor expressed as the
integral of the limb-darkening weighted by the Legendre polynomial
with the $\ell$ degree. Derivatives of the monochromatic flux,
${\cal F}_\lambda(T_{\rm eff},\log g)$, are calculated from static
atmosphere models. In general, they depend also on the metallicity
parameter [m/H] and microturbulence velocity $\xi_t$. Here, we use
the Kurucz (2004) atmosphere models and Claret (2000) limb-darkening
coefficients.

As was already mentioned in Introduction, the $f$ parameter
describes the ratio of the bolometric flux perturbation to the
radial displacement at the level of the photosphere:
$$\frac{ \delta {\cal F}_{\rm bol} } { {\cal F}_{\rm bol} }=
{\rm Re}\{ \varepsilon f Y_\ell^m(\theta,\varphi) {\rm e}^{-{\rm i}
\omega t} \}.\eqno(3)$$
The value of $f$ can be obtained from linear computations of stellar
pulsation; because of nonadiabatic character of the pulsation, this
quantity is complex. Here we use the pulsational code of W. Dziembowski.

If data on radial velocity variations are available, then the
amplitude and phases of these variations can be included in a
process of mode identification. The radial velocity is determined
from observations as the first moment of well separated line
profile, ${\cal M}_1^{\lambda}$. According to Dziembowski (1977),
the expression for the complex amplitude of the radial velocity
variations is the following
$${\cal A}_{\rm Vrad}= {\cal M}_1^{\lambda}= \varepsilon Y_{\ell}^m(i,0) {\rm i}\omega R \left( u_{\ell}^{\lambda}
+ \frac{GM}{\omega^2R^3} v_{\ell}^{\lambda} \right),\eqno(4)$$
where $u_{\ell}^{\lambda}$, $v_{\ell}^{\lambda}$ are another
disc-averaging factors introduced by Dziembowski (1977).

In principle, two approaches can be used in mode identification. The
first approach makes use of amplitude ratios and phases differences
and employs input from the pulsation theory, i.e., the $f$
parameter. In the linear and non-rotation approximation, the
amplitude ratios and phases differences are independent of the
intrinsic mode amplitude, $\varepsilon$, azimuthal order, $m$, and
the inclination angle, $i$, because the factor $\varepsilon
Y_\ell^m(i,0)$ cancels out.

The second method employs amplitudes and phases alone in order
to determine from data the empirical values of $f$ and the intrinsic
amplitude dependent on inclination, $\tilde\varepsilon=\varepsilon
Y_\ell^m(i,0)$, together with the mode degree, $\ell$.

Both methods require input from models of stellar atmospheres.

\subsection{Determination of the $\ell$ degree using photometry and theoretical values of $f$}

\begin{figure}
\centering
\includegraphics[width=88mm,clip]{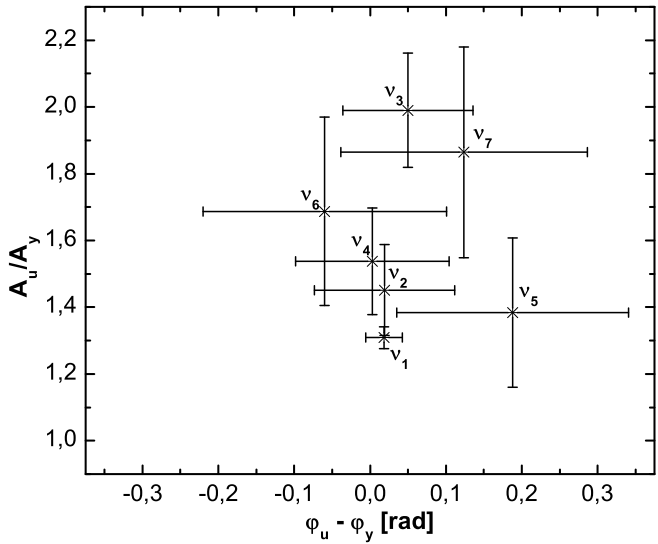}
\caption{ Photometric diagnostic diagram for the $uy$ Str\"omgren passbands
with positions of seven pulsational frequencies detected by Handler et al. 2005.}
\label{aaaaaa}
\end{figure}
\begin{figure*}
\centering
\includegraphics[width=17.5cm,clip]{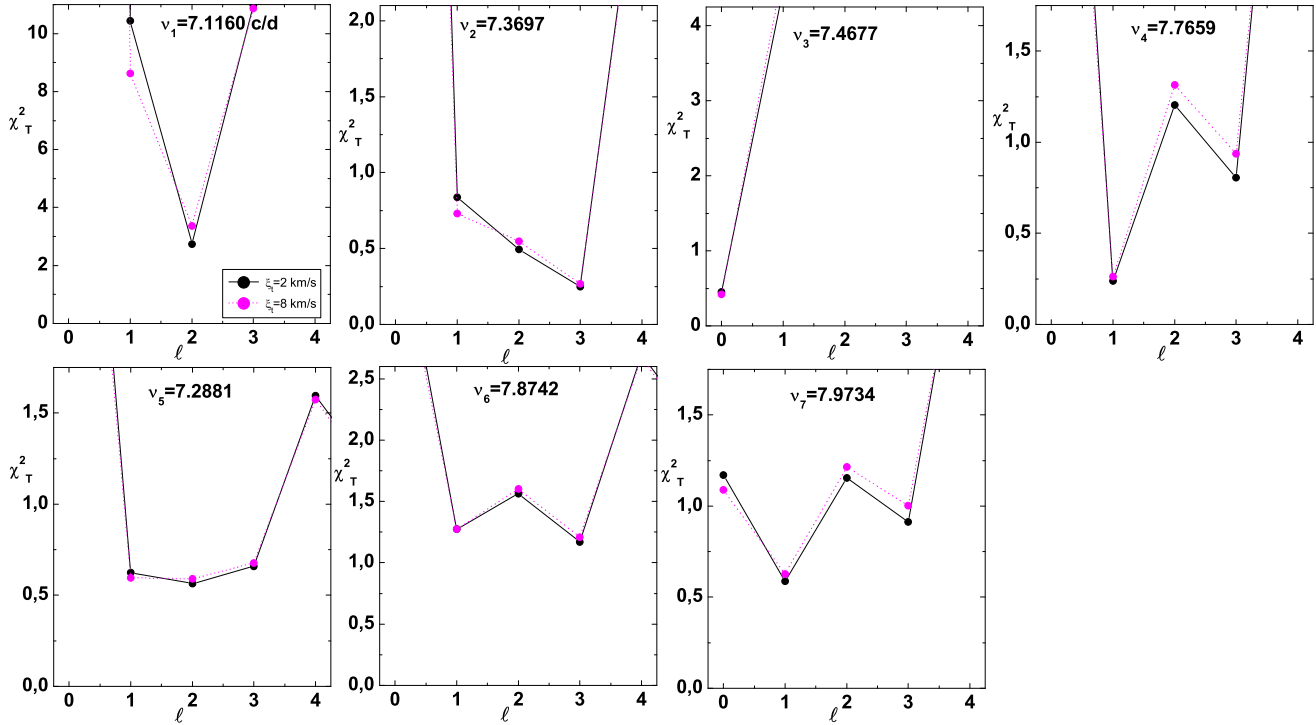}
\caption{Identification of the $\ell$ degrees of the pulsational
modes detected in the photometric variations using amplitude ratios
and phase differences and theoretical $f$ parameter
($\chi^2_T~vs.~\ell$). These results correspond to the model with
the following stellar parameters: $M=9.5 M_\odot,~\log T_{\rm
eff}=4.363,~ Z=0.015$, and computed with the OP-A4 opacity. The
Kurucz atmosphere models were adopted. The effect of the
microturbulence velocity, $\xi_t$, on the determination of $\ell$ is
shown. } \label{aaaaaa}
\end{figure*}

Observations of $\theta$ Oph in the $uvy$ Str\"omgren photometry
were kindly provided by Gerald Handler. From these data we
determined amplitudes and phases of the seven frequencies detected
by HSM05 and, following HSM05, the amplitudes were corrected for the
light contribution from the speckle companion:
$A_u^{cor}=1.19A_u,~A_v^{cor}=1.22A_v,~A_y^{cor}=1.23A_y$. Although
the phase differences between photometric passbands are small, they
are not negligible, as shown in Fig.\,3 for the $uy$ pair.
Therefore, we included also these quantities to discriminate between
various $\ell$ degrees. To this end we use, as usually, the
following discriminant:
$$\chi^2_T=\frac1{2(N-1)} \sum_{i=1}^{N-1} \frac{(X_i^o-X_i^t)^2}{ \sigma_i^2}, \eqno(5)$$
where $X_i^o,~X_i^t$ are observational and theoretical photometric
observables, respectively, such as $A_{\lambda_i}/A_{\lambda_1}$ and
$\varphi_{\lambda_i}-\varphi_{\lambda_1}$. N is the number of
passbands and if we use only amplitude ratios we have to skip 2 in
the denominator before the sum. The index ${\lambda_1}$ denotes the
photometric passband to which the others, ${\lambda_i}$, are
normalized and $\sigma_i$ are the observational errors of the
$X_i^o$ quantities. Because we treat measurements in different
passbands as independent variables, the formulae for errors
($\sigma_i$) are:
$$\sigma^2\left(\frac{A_{\lambda_i}}{A_{\lambda_1}}\right)= \left(\frac{A_{\lambda_i}}{A_{\lambda_1}}\right)^2
\left[ \frac{ \sigma^2(A_{\lambda_i})} {A^2_{\lambda_i}} +
       \frac{ \sigma^2(A_{\lambda_1})} {A^2_{\lambda_1}} \right], \eqno(6)$$
%\left[ \left( \frac{ \sigma(A_{\lambda_i})} {A_{\lambda_i}} \right)^2 +
%       \left( \frac{ \sigma(A_{\lambda_1})} {A_{\lambda_1}} \right)^2 \right]$$
%
and
$$\sigma^2(\varphi_{\lambda_i}-\varphi_{\lambda_1})= \sigma^2(\varphi_{\lambda_i}) + \sigma^2(\varphi_{\lambda_1}). \eqno(7)$$
We perform mode identification for the whole range of allowed stellar parameters and
different values of metallicity.

In Fig.\,4, we show results of our mode identification for the seven
pulsational frequencies of $\theta$ Oph. This is an example for the
model with parameters: $M=9.5 M_\odot,~\log T_{\rm
eff}=4.363,~X=0.7,~ Z=0.015$, computed with the OP-A04 opacities.
The character of $\chi^2_T(\ell)$ for the other models is
qualitatively the same. We consider stellar atmospheres with two
values of the microturbulence velocity: $\xi_t=2$ and 8 km/s.
{\bf In some cases a value of $\chi^2_T$ is less than 1
which comes from large observational errors.}

Our identifications of the $\ell$ degrees agree with those obtained
by HSM05 who used only photometric amplitudes. A discussion of the
$\ell$ identification for each frequency will be given at the end of
the next subsection.

%\subsection{Determination of the $\ell$ degree from combined photometry and spectroscopy, using empirical $f$ values}
\subsection{Determination of the $\ell$ degree using empirical values of $f$}
Three pulsational frequencies ($\nu_1,~\nu_2, ~\nu_3$) were found in
both photometric and spectroscopic observations of $\theta$ Oph. By
combining these two types of data, we can identify the $\ell$ degree
of these three frequencies and extract corresponding empirical
values of the nonadiabatic $f$ parameter. In our analysis we use the
radial  velocity as determined by the first moment of the SiIII4553
spectral line. These data were kindly provided by Maryline Briquet.
In our analysis we included only these spectra which coincide in
time photometric observations.
\begin{figure*}
\centering
\includegraphics[width=17.5cm,clip]{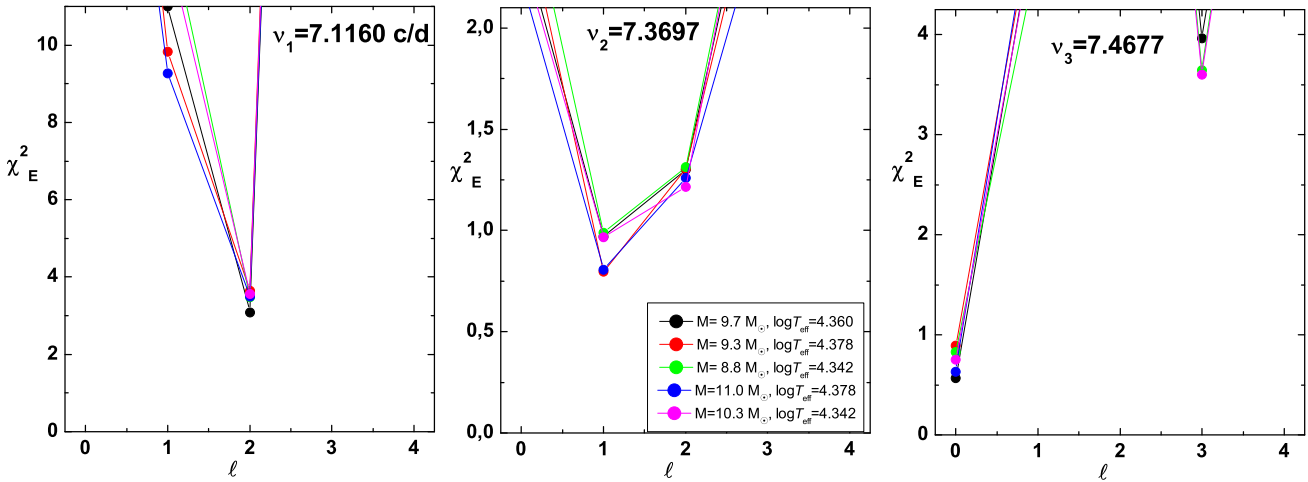}
\caption{Identification of the $\ell$ degrees of the pulsational
modes detected in both photometry and in spectroscopy
($\chi^2_E~vs.~\ell$). Various lines correspond to uncertainties in
the stellar parameters (see text for more explanation). The Kurucz
atmosphere models with the microturbulence velocity of $\xi=2$ km/s
were adopted. } \label{aaaaaa}
\end{figure*}

The method of simultaneous determination of $\ell$ and $f$ has been
described in details by DD03 and DD05, and we will not repeat this
here. Each passband, $\lambda$, yields l.h.s. of Equations (1)
whereas data on the radial velocity yield l.h.s. of Equation (4).
This set of linear equations is solved for a specified $\ell$ in order
to determine two complex {\bf unknown} quantities: $\tilde\varepsilon$ and
$\tilde\varepsilon f$. With three passbands and the radial velocity
data we have a set of four complex equations. The idea is to derive
the value of the nonadiabatic parameter $f$ and the intrinsic
amplitude $\tilde\varepsilon$ from observations, which fit best the
theoretical and observational amplitudes and phases for a given
$\ell$ degree. The goodness of fit is calculated as
$$\chi^2_E=\frac1{2N-M} \sum_{i=1}^N
 \frac{ |{\cal A}^o_{\lambda_i} - {\cal A}^t_{\lambda_i}|^2 }{ |\sigma_{\lambda_i}|^2}, \eqno(8)$$
where $N$ is the number of passbands plus one (to account for the
fact that we use the radial velocity measurements) and $M$ is the
number of parameters to be determined. Here $\lambda$ denotes either
the photometric passband or radial velocity. ${\cal A}^o$ and ${\cal
A}^t$ are complex observational and theoretical amplitudes,
respectively, and $\sigma_{\lambda_i}$ are the observational errors.
Because we treat amplitudes and phases as independent variables, the
formula for the errors is
$$|\sigma_\lambda|^2= \sigma^2 (A_{\lambda})  +  A_{\lambda}^2 \sigma^2(\varphi_\lambda), \eqno(9) $$
where $A_{\lambda}=|{\cal A}_\lambda|$ and $\varphi_{\lambda}=arg({\cal A}_\lambda)$.

The results of mode identification by the above method for the first
three frequencies of $\theta$ Oph are shown in Fig.\,5. Various
lines correspond to the uncertainties in the stellar parameters. We
calculated $\chi^2_E$ for the models from the center and from the
four edges of the error box in Fig.\,1.

In Table\,1 we give a summary of the $\ell$ identification from the
two approaches. From both methods, the dominant frequency is
undoubtedly a $\ell=2$ mode and the $\nu_3$ frequency is a radial
mode. In the case of the $\nu_2$ frequency, we were able to exclude
the $\ell=3$ degree from combined multicolour photometry and radial
velocity data. For the remaining frequencies, we can rely only on
mode identification from photometry alone. The three frequency
peaks, $\nu_4,~\nu_6$ and $\nu_7$, are equidistance and most
probably form a close $\ell=1$ triplet. Other values of $\ell$ are
also possible but because the disc averaging effects increase very
rapidly with $\ell$, we regard these higher degrees as much less
probable and assume, as HSM05, that this is the $\ell=1$ triplet.
The $\ell=0$ identification for $\nu_7$ is of course excluded
because $\nu_7$ is too close to $\nu_3$ which is the radial mode.
The $\nu_5$ frequency can be identified either as $\ell$=1 or 2 or
3, with a lower probability for the $\ell=3$ identification. For the
dominant mode, B05 identified also the azimuthal order, $m=-1$, on
the basis of amplitudes and phases across the SiIII line profile.

Although there are some regularities in frequency spacing between
$\nu_1,~\nu_5$ and $\nu_2$, we are more cautious about believing
that they are components of the same quintuplet. The regularities
suggest that $\nu_5$ may be the $(\ell,m)=(2,+1)$ mode and $\nu_2$ -
the $(\ell,m)=(2,+2)$ mode, as was assumed by Briquet et al. (2007)
in their seismic analysis of $\theta$ Oph. However, from the
amplitude and phase diagrams computed by B05 from variations of the
SiIII line profile, the frequency $\nu_3=7.3697$ c/d looks rather
like a $m=0$ mode, because the phase change is close to $\pi$ around
the line center (see Fig.\,3 of B05).

Thus, our complex seismic analysis we will be based on frequencies
$\nu_3$ and $\nu_6$ which are well identified centroid modes with
$\ell=0$ and $\ell=1$, respectively, and we postpone studies of
rotational effect until more than one multiplet will be well
identified.
\begin{table}
\caption{Summary of mode identification using two methods for frequency peaks detected in $\theta$ Ophiuchi.}
\begin{tabular}{|c|c|c|}
\hline
   Frequency [c/d] &         photometry         & photometry+$V_{\rm rad}$ \\
                   &  theoretical $f$ parameter & empirical $f$ parameter  \\
\hline
   $\nu_1$= 7.11600  & $\mathbf{\ell=2}$  & $\mathbf{\ell=2}$ \\
\hline
   $\nu_2$= 7.3697   & $\mathbf{\ell=3,2,1}$  & $\mathbf{\ell=1,2}$ \\
\hline
   $\nu_3$= 7.4677   & $\mathbf{\ell=0}$  & $\mathbf{\ell=0}$ \\
\hline
   $\nu_4$= 7.7659   & $\mathbf{\ell=1,3}$  & \textbf{--} \\
\hline
   $\nu_5$= 7.2881   & $\mathbf{\ell=1,2,3}$  & -- \\
\hline
   $\nu_6$= 7.8742   & $\mathbf{\ell=1,3,2}$  & -- \\
\hline
   $\nu_7$= 7.9734  & $\mathbf{\ell=1,3,2,0}$  & -- \\
\hline
\end{tabular}
\end{table}

%\section{Inferences from the $f$ parameter of the $\ell=0$ mode}
\subsection{Determination of the radial orders}
In Fig.\,1 we have plotted lines of a constant period (0.1339 d) for
fundamental ($n=1$) and first overtone ($n=2$) radial mode. These
line correspond to $\nu_3=7.4677$ c/d.  As we can see, within the
allowed space of stellar parameters there are two possible
identifications for the radial order: $n=1$ or $n=2$. To
discriminate between these two identifications, we shall use results
from the method of DD03 (Section 3.2).

Firstly, we can calculate values of $\chi^2_E$ as a function of
effective temperature for the case of $n=1$ and $n=2$. This is shown
in Fig.\,6. The vertical lines mark the allowed range of $\log
T_{\rm eff}$. Although the fundamental mode takes lower values of
$\chi^2_E$, we cannot exclude the first overtone from this graph
because both lines reach acceptable values of $\chi^2_E$.

Another way to discriminate between the two radial orders for the
$\ell=0$ mode is to compare theoretical and empirical values of the
nonadiabatic $f$ parameter (see DD05). The photometric and
spectroscopic data for $\theta$ Oph allowed to determine the values
of $f$ for the $\nu_3$ mode with sufficient accuracy to use them to
this aim. In Fig.\,7 we show such comparison. Only models with
stellar parameters within the error box of Fig.\,1 were considered.
As we can see, the $n$ identification is clear and $\nu_3$ is the
radial fundamental mode. In addition, lines of constant values of
the instability parameter, $\eta=0$, are shown. Only models to the right
of this line have unstable radial modes.
In Fig.\,7, we show also the sensitivity of
the theoretical $f$ parameter to metallicity prameter $Z$. A
comparison of theoretical and empirical values of $f$ gives also
additional support to the result that metallicity of $\theta$ Oph is
lower than $Z=0.015$.
\begin{figure}
\centering
\includegraphics[width=88mm,clip]{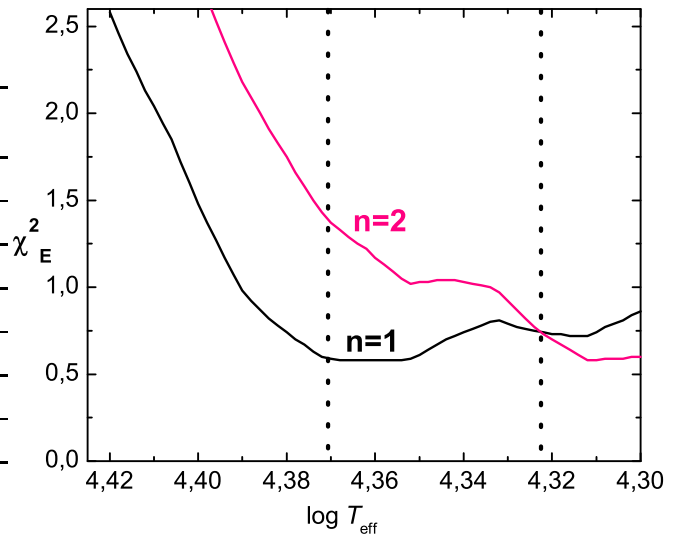}
\caption{ The $\chi^2_E$ as a function of the effective temperature
for the frequency $\nu_3$ identified as the l=0 mode. Two possible
values of the radial order, $n$, were considered. Vertical dotted
lines indicate the allowed range of $\log T_{\rm eff}$.}
\label{aaaaaa}
\end{figure}
\begin{figure}
\centering
\includegraphics[width=88mm,clip]{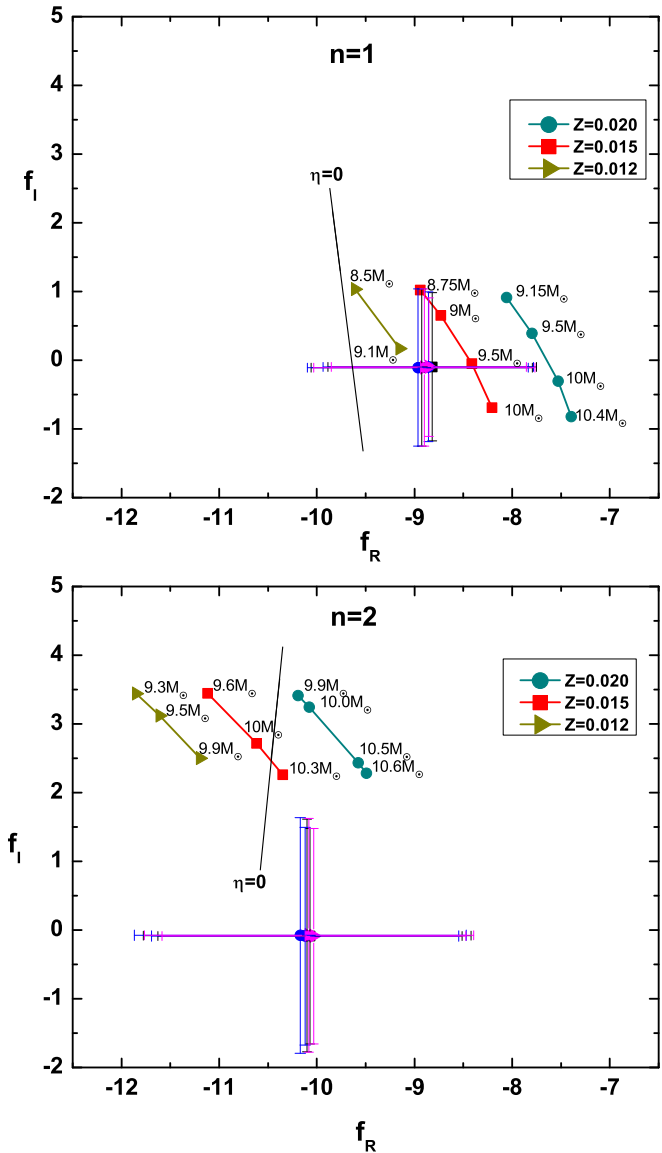}
\caption{Comparison of empirical and theoretical values of the $f$
parameter for the radial mode $\nu_3$. The upper panel refers to the
fundamental mode hypothesis and the lower one, to the first overtone
mode hypothesis. Theoretical values of $f$ were calculated with the
OP-A04 opacities, metallicity $Z=0.020$, 0.015 and 0.012, and
assuming stellar parameters within the error box. Lines of constant values
of the instability parameter, $\eta$, are labeled as $\eta=0$.}
\label{aaaaaa}
\end{figure}

After fixing the $n$ order of the $\nu_3$ radial mode as $p_1$, we
can try to associate the radial orders also for other frequencies. A
survey of pulsational models showed that the $\ell=1$ triplet can be
identified as $p_1$, whereas the dominant $(\ell,m)=(2,-1)$ mode, as
a component of the $g_1$ quintuplet.

\section{Constraints on intrinsic mode amplitudes}
Having determined the value of $\tilde\varepsilon=\varepsilon
Y_\ell^m(i,0)$ from the method of DD03 in Section 3.2, we can get a
lower limit for the value of the intrinsic mode amplitude,
$\varepsilon$, and, in the case of a radial mode, its exact value.
The available observations of $\theta$ Oph allow to determine the
values of $\tilde\varepsilon$ for $\nu_1$ and $\nu_3$ with a
sufficient accuracy.

The dominant  mode, $\nu_1$, is unequivocally the $\ell=2$ mode.
Moreover, from spectroscopic time series analysis B05 identified its
azimuthal order as $m=-1$. With these angular numbers, we can
calculate the intrinsic amplitude of the dominant mode as a function
of the inclination angle. The result is shown in Fig.\,8. As one can
see, inclination angles close to $85-90^\circ$ are excluded because
of the node line of the $(\ell,m)=(2,-1)$ mode. The minimum value of
the intrinsic amplitude of the dominant mode is equal to 0.0098 at
i=45$^\circ$, i.e. 0.98\% of the stellar radius. In the case of the
$\nu_3$ frequency, which correspond to the radial mode, we estimated
the exact value of $\varepsilon$ as 0.0014, i.e. 0.14\% of the
stellar radius. These estimates mean that the intrinsic amplitude of
the dominant $(\ell,m)=(2,-1)$ mode is at least 7 times larger than
that of the radial mode. Here the open problem of mode selection
emerges and such determinations can bring us closer to a solution.
\begin{figure}
\centering
\includegraphics[width=88mm,clip]{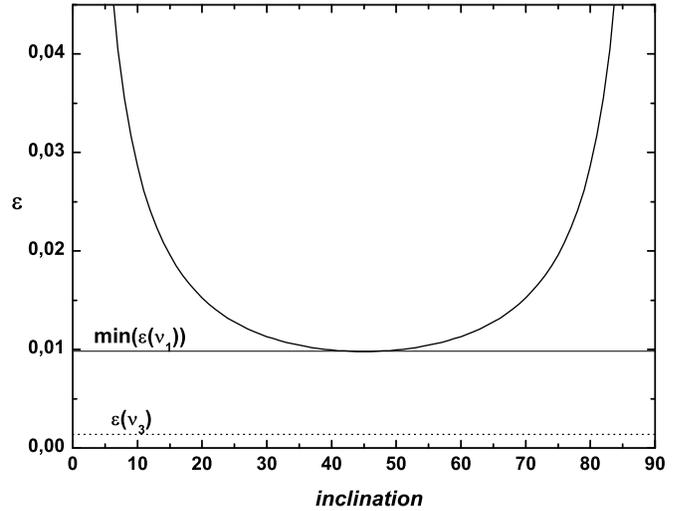}
\caption{The intrinsic amplitude of the dominant  mode, $\nu_1$,
$(\ell,m)=(2,-1)$, as a function of the inclination angle. The
minimum value of $\varepsilon(\nu_1)$ is equal to about 0.01. The value of
$\varepsilon$ for the radial mode, $\nu_3$, is plotted as the
dotted line.} \label{aaaaaa}
\end{figure}

\section{Complex asteroseismic modelling}
We shall perform our seismic modelling in two steps. Firstly, we
find a family of models which fit the two centroid frequencies of
$\theta$ Ophuichi, i.e., $\nu_3$, which was identified as the
$\ell=0,p_1$ mode, and $\nu_6$, which is the centroid of the
$\ell=1,p_1$ triplet. In the next step, from this set of models we
will select only those which reproduce, within the observational
errors, the empirical values of $f$ of the radial fundamental mode.

\subsection{Fitting frequencies of the $\ell=0$ mode and the $\ell=1$ centroid mode.}
To construct our seismic models, we use two frequencies:
$\nu_3=7,4677$ (the $\ell=0$ mode) and $\nu_6=7.8742$ c/d (the
$\ell=1$ centroid mode), because the $\nu_1$ frequency is a
non-axisymmetric $\ell=2$ mode with $m=-1$, while frequencies
$\nu_2$ and $\nu_5$ do not have unambiguous identification. We
searched for models fitting these two frequencies by changing
various parameters: the mass, $M$, the effective temperature,
$T_{\rm eff}$, the chemical abundance, ($X,Z$), and the core
overshooting parameter, $\alpha_{\rm ov}$. The effect of
overshooting was computed according to the new formulation of
Dziembowski \& Pamyatnykh (2008). This new treatment takes into
account not only the distance of the overshooting but also partial
mixing in the overshoot layer. In our seismic modelling, we relied
on the latest determination of the solar element mixture by A04 and
used both the OP tables (Seaton 2005) and OPAL tables (Iglesias \&
Rogers 1996). The rotational splitting for the $\ell=1$ triplet is
about 0.1 c/d, giving the rotational velocity of $V_{\rm rot}=30$
km/s for the stellar radius of about $R=6 R_{\odot}$; this is
consistent with the $V_{\rm rot}\sin i$ value derived by Abt, Levato
\& Grosso (2002) mentioned in Section 2. We assumed this value of
$V_{\rm rot}$ in all computations.

In Fig.9 we show the results of our seismic modelling for the OP
data on the $\alpha_{\rm ov}~vs.~Z$ plane. The instability borders
are labeled with $\eta=0$ and shown as the thick solid and thick
dotted lines for the $\ell=0$ and $\ell=1$ modes, respectively. Only
models above those lines are unstable and reproduce the observed
frequencies $\nu_3$ and $\nu_6$. Moreover, we depicted also the
lines of constant mass and effective temperature. We show results
for two values of hydrogen abundance: $X=0.7$ (the left panel) and
$X=0.75$ (the right panel).

From such data, one can easily find the relation for a family of
seismic models for a fixed value of $X$ in the form:
$$\alpha_{\rm ov}=a\cdot M+b\cdot\log T_{\rm eff} + c\cdot Z +d.$$
As we can see, models with lower $Z$ require more core overshooting.
This is because for a higher value of $Z$ the convective core is
larger, implying a lower value of the overshooting distance. The
same result was obtained by Briquet et al. (2007) for $\theta$ Oph
and by Dziembowski \& Pamyatnykh (2008) in their studies of hybrid
pulsators, $\nu$ Eri and 12 Lac. A similar dependence was also
obtained for lower mass stars (M.-J. Goupil, private communication).
On the other hand, at a given metallicity, for a higher value of $X$
we need higher value of $\alpha_{\rm ov}$ to get pulsational
instability.

All these seismic models of $\theta$ Oph prefers lower effective
temperatures and lower masses. The lower effective temperature limit
is $\log T_{\rm eff}\approx 4.34$, thus most of these seismic models
are outside the error box. We would like to mention that a lower
value of effective temperature was determined from the IUE spectra
by Niemczura\& Daszy\'nska (2005) who derived $\log T_{\rm
eff}=4.347\pm 0.016$. The ultraviolet spectra of the $\theta$ Oph
system are dominated by the main component, i.e. the early B-type
star, which emits most energy in UV and dominates this part of
spectrum. Therefore, this lower value of $\log T_{\rm eff}$ can be
more reliable and it is also more consistent with seismic models.
The higher value of $X$ require even lower values of $\log T_{\rm
eff}$. The effect of the opacity tables will be discussed in the
next section.
\begin{figure*}
\centering
\includegraphics[width=17.5cm,clip]{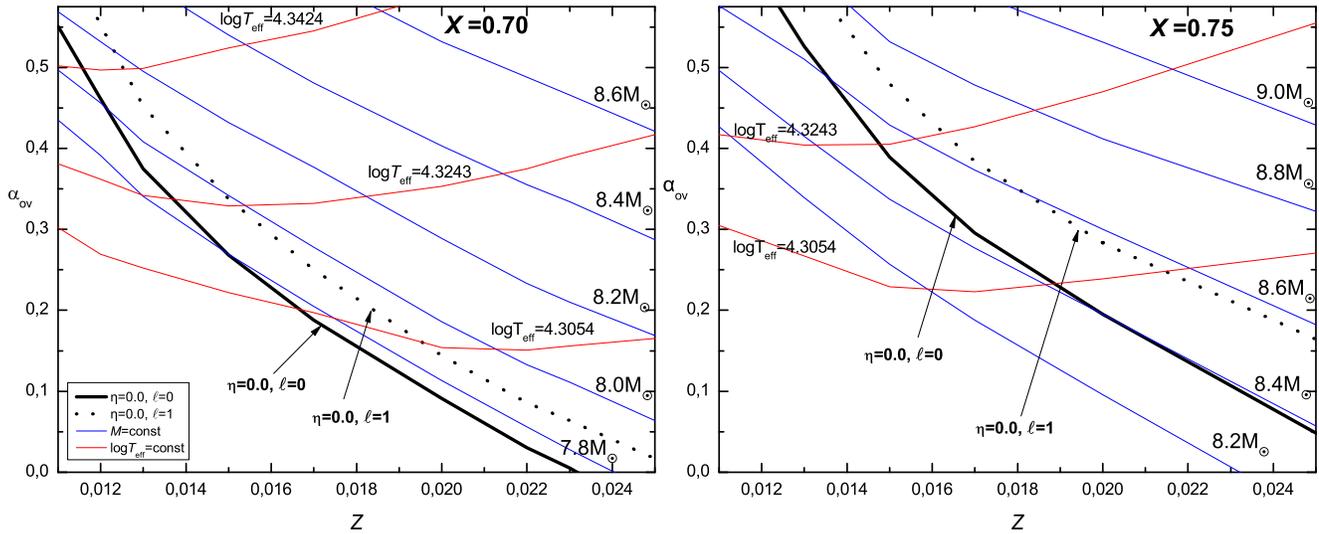}
\caption{The overshooting parameter as a function of metallicity for
the seismic models of $\theta$ Oph found from the fitting of the
$\nu_3$ frequency (the $\ell=0,p_1$ mode) and the $\nu_6$ frequency
(the centroid of the $\ell=1,p_1$ triplet). Results for two values
of the hydrogen abundance, $X$, are shown: 0.70 (the left panel) and
0.75 (the right panel). } \label{aaaaaa}
\end{figure*}

\subsection{Fitting the $f$ parameter of the $\ell=0, p_1$ mode}
Determination of empirical values of $f$ puts additional and unique
constraints on seismic models. In that way, we are going one step
further by a requirement of fitting simultaneously pulsational
frequencies and the corresponding values of $f$. The sensitivity of
the theoretical $f$ parameter to metallicity is very strong and was
already shown in Fig.\,7. How this quantity depends on the hydrogen
abundance, overshooting distance and adopted opacity tables is
demonstrated in Fig.\,10. As we can see, the strongest effect comes
from opacity data. The amount of hydrogen and the overshooting
distance have a minor effect on the value of $f$.

From a set of seismic models found in Section 5.1, we singled out
only those which have the $f$ parameter for the radial mode
consistent with the observed values within the errors. Having also
another source of opacity data, we studied the effect of this input
on the seismic models. In Fig.\,11 we plotted similar diagrams as in
Fig.\,9, but at a fixed hydrogen abundance ($X=0.7$) and considering
both the OP tables (the left panel) and the OPAL tables (the right
panel). Moreover, we superimposed areas resulting from fitting
empirical values of $f$ for the radial mode. As we can see, only
models calculated with the OPAL tables can account for the empirical
$f$ parameters with a reasonable value of the overshooting distance.
An interesting fact is that with both opacity tables it was possible
to reproduce the real parts of $f$ for lower values of $\alpha_{\rm
ov}$, whereas only the OPAL data gave consistency also in the
imaginary parts of $f$. This is a very important and revealing
result. It shows how complex asteroseismic studies can test data on
microscopic physics. Another support for the OPAL data is that with
these opacities we get also a better agreement between stellar
parameters of seismic models and those derived from photometric or
spectroscopic observations.
\begin{figure}
\centering
\includegraphics[width=88mm,clip]{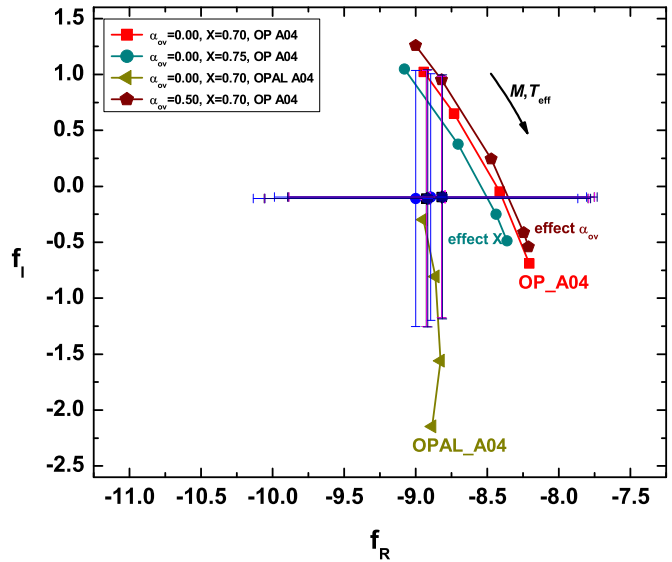}
\caption{A comparison of empirical and theoretical values of $f$ for
the radial mode of $\theta$ Oph. Theoretical $f$ parameters were
computed for different values of the hydrogen abundance, $X$, the
overshooting parameter, $\alpha_{\rm ov}$, and for two sources of
opacity data. The A04 solar chemical mixture was adopted and
metallicity fixed at Z=0.015. } \label{aaaaaa}
\end{figure}

\begin{figure*}
\centering
\includegraphics[width=17.5cm,clip]{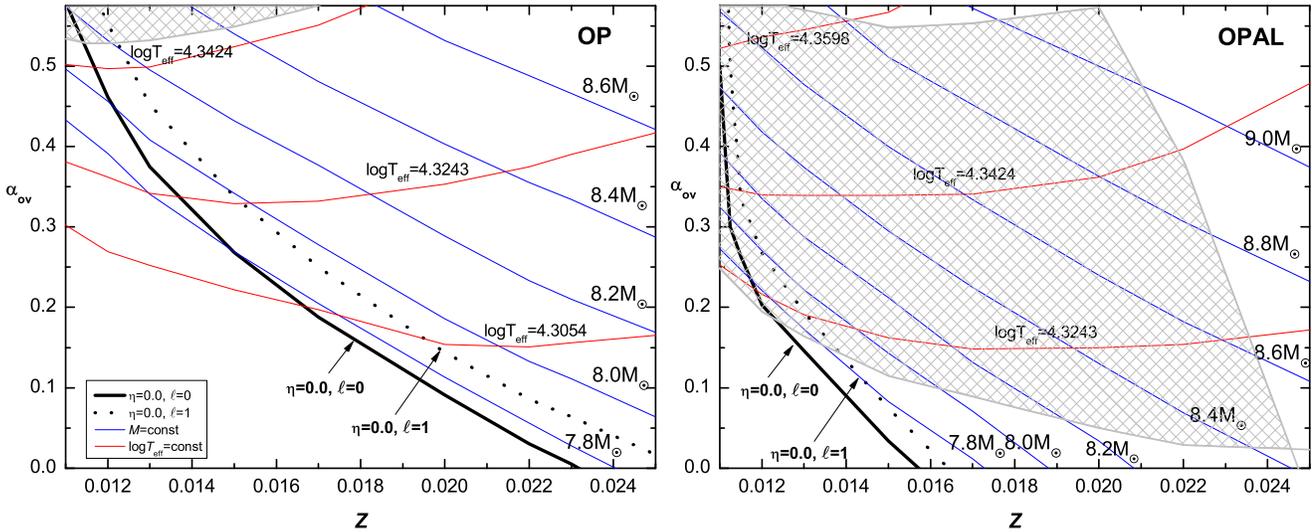}
\caption{The overshooting parameter as a function of metallicity for
the seismic models of $\theta$ Oph found from the fitting of the
$\nu_3$ frequency (the $\ell=0,p_1$ mode) and the $\nu_6$ frequency
(the centroid of the $\ell=1,p_1$ mode). Results were obtained with
the OP (the left panel) and the OPAL table (the right panel). The
hatched areas mark the models which fit the $f$ parameter for the
radial fundamental mode. } \label{aaaaaa}
\end{figure*}

\section{Summary}
Our analysis showed a great potential of probing parameters of
stellar models and microphysics by means of the nonadiabatic
parameter $f$. Incorporating this quantity in seismic modelling
provides much stronger constraints and a new kind of information.
The adequate seismic model should account not only for oscillation
frequencies but also for the empirical values of $f$, visibility and
instability conditions. This kind of seismic modelling we called
{\it complex asteroseismology}. The aim of the complex asteroseismic
analysis is to find stellar models which fit both oscillation
frequencies and corresponding $f$ parameters. The value of $f$ is
determined in subphotospheric layers which only weakly affect the
pulsational frequencies. Therefore, these two asteroseismic tools
($\nu,f$) are complementary to each other and treating them
simultaneously can improve seismic modelling of any kind of
pulsating stars. As was already discussed by DD05, for B-type
pulsators, the $f$ seismic tool probes in particular stellar
metallicity and opacities. In this paper, we have demonstrated that
even with data on only two well identified centroid modes and the
empirical $f$ parameter for the radial mode we can derive very
strong constraints on stellar parameters, core overshooting and, in
particular, on opacities.

Our analysis of $\theta$ Ophiuchi showed that only with the OPAL
table it was possible to reproduce simultaneously the two
pulsational frequencies and the empirical value of $f$ corresponding
to the radial mode. Seismic models obtained with the OP data require
enormous amounts of core overshooting, at least $\alpha_{\rm
ov}=0.55$, which we consider not physical for a mass and rotational
velocity appropriate for $\theta$ Oph. Computations with the OPAL
tables allowed to achieve agreement between observed and theoretical
values of $f$ for a wide space of parameters, $(M, T_{\rm
eff},\alpha_{\rm ov}, Z)$, fixed by the mode frequency fitting. Such
complex asteroseismic studies provide a critical and unique test for
stellar opacities and the atomic physics.

Another important results of this paper are valuable constraints on
the intrinsic amplitude for the dominant $(\ell=2,m=-1)$ mode and
for the $\ell=0$ mode. We have found out that the dominant mode of
$\theta$ Oph has to be excited with the intrinsic amplitude at least
seven times larger than the radial mode. Although, in the framework
of linear theory, we cannot compare these values with theoretical
predictions, such estimates can give us some guidelines to the mode
selection mechanism in the $\beta$ Cephei pulsators.

\section*{Acknowledgments}
We gratefully thank Gerald Handler and Maryline Briquet for kindly
providing photometric observations and data on moments of the
SiIII4553 spectral line of $\theta$ Oph, respectively. We appreciate
also discussion with Gerald Handler on observational aspects and
thank for his useful comments. We are very indebted
to Mike Jerzykiewicz for carefully reading the manuscript.
These results were also presented during the COROT2009/HELASIII symposium
held in Paris, February, 2-5, 2009. A participation of JDD at this conference
was financially supported by the EU HELAS Network, 6PR, No. 026138.
%The work was supported by the Polish MNiSW grant No. ...

\label{lastpage}

\end{document}